\documentclass[prd,twocolumn,amsmath,amssymb,floatfix,superscriptaddress]{revtex4-1}

\usepackage{graphicx}
\usepackage{amssymb}
\usepackage{amsmath}
\usepackage{hyperref}

\usepackage{color}

\DeclareMathAlphabet\mathbfcal{OMS}{cmsy}{b}{n}
\definecolor{darkgreen}{cmyk}{0.85,0.2,1.00,0.2} 
\definecolor{purple}{cmyk}{0.5,1.0,0,0} 
\def\physrep{Phys.~Rep.}

\newcommand{\wh}[1]{{#1}}

\newcommand{\hidden}[1]{}

\def\barray{\begin{array}} 
\def\earray{\end{array}}
\def\be{\begin{equation}}
\def\ee{\end{equation}}
\def\ben{\begin{equation} \nonumber}
\def\een{\end{equation}}
\def\ban{\begin{eqnarray*}}
\def\ean{\end{eqnarray*}}
\def\ba{\begin{eqnarray}}
\def\ea{\end{eqnarray}}

\def\({\left(}
\def\){\right)}

\newcommand{\simgt}{\lower.5ex\hbox{$\; \buildrel > \over \sim \;$}}
\newcommand{\simlt}{\lower.5ex\hbox{$\; \buildrel < \over \sim \;$}}

\newcommand{\mnras}{Mon.~Not.~Roy.~Astron.~Soc.}

\newcommand{\bx}{{\bf x}}
\newcommand{\bk}{{\bf k}}

\newcommand{\bq}{{\bf q}}

\newcommand{\lin}{{\rm L}}

\newcommand{\tW}{\tilde{W}}
\newcommand{\tu}{\tilde{u}}
\newcommand{\fpole}{{\cal F}}

\begin{document}

\title{Super-Sample Covariance in Simulations}
\author{Yin Li}
\affiliation{Department of Physics, University of Chicago, Chicago, Illinois 60637, U.S.A.}
\affiliation{Kavli Institute for Cosmological Physics, Department of Astronomy \& Astrophysics,  Enrico Fermi Institute, University of Chicago, Chicago, Illinois 60637, U.S.A.}
\author{Wayne Hu}
\affiliation{Kavli Institute for Cosmological Physics, Department of Astronomy \& Astrophysics,  Enrico Fermi Institute, University of Chicago, Chicago, Illinois 60637, U.S.A.}
\author{Masahiro Takada}
\affiliation{Kavli Institute for the Physics and Mathematics of the Universe
(Kavli IPMU, WPI),
The University of Tokyo, Chiba 277-8583, Japan}
\begin{abstract}
Using separate universe simulations, we accurately quantify super-sample covariance (SSC), the typically dominant sampling error for
  matter power spectrum estimators in a finite volume, which  arises from the presence of super survey modes.  {By quantifying the power spectrum response to a background mode}, this approach automatically captures the separate effects of beat
coupling in the quasilinear regime, halo sample variance in the nonlinear regime and
a new dilation effect which changes scales in the power spectrum coherently across
the survey volume, including the baryon acoustic oscillation scale.   It models these effects at typically the few percent level or better with a handful of small
volume simulations for any survey geometry compared  with directly using
many thousands of survey volumes in a suite of large volume simulations.  The stochasticity of
the response is sufficiently small that in the quasilinear regime, SSC can be alternately
included by fitting the mean density in the volume with these fixed templates in parameter estimation.
{
We also test the halo model prescription and find agreement typically at better than the 10\% level for the response.}   
 \end{abstract}

\maketitle
\section{Introduction}

The statistical properties of large scale structure provide a wealth of cosmological
information on fundamental physics, including cosmic acceleration, neutrino masses and inflation.   The simplest
statistic  is the two-point correlation function or power spectrum of the  matter density field which underlies observables such as weak lensing and galaxy clustering.  
Its covariance matrix encapsulates the precision with which it can be measured and
contains contributions from both
measurement noise and  
covariances caused by the incomplete sampling of the fluctuations due to a
finite-volume survey \cite{MeiksinWhite:99,Scoccimarroetal:99,WhiteHu:00}. 

 Super-sample covariance (SSC) is the sampling error caused
by  coupling to modes that are larger than the survey scale \cite{Hamiltonetal:06}.  It has been shown to be
the largest non-Gaussian contribution to power spectrum errors for a wide
range of conditions from the quasilinear regime, where it is known as beat coupling,
to the deeply nonlinear regime, where it is known as halo sample variance
\cite{Hamiltonetal:06,Sefusattietal:06,HuKravtsov:03,TakadaBridle:07,TakadaJain:09,Satoetal:09,Takahashietal:09,dePutter:2011ah,Schneider:2011wf,Kayoetal:13,TakadaSpergel:13,Takada:2013wfa}.  

Accurate quantification of SSC is therefore crucial for upcoming surveys.   In Ref.~\cite{Takada:2013wfa}, it was shown that all previously known aspects of SSC could
be characterized by a single quantity: the response of the power spectrum to a change
in the background density.   Indeed  
by following this prescription we here uncover a hitherto uncharacterized dilation effect in the quasilinear
regime and discuss its origin in the matter trispectrum.

This power spectrum response can itself be accurately calibrated with
so-called separate universe simulations, where the long-wavelength 
perturbation is absorbed into a change in cosmological parameters
\cite{Sirko:05,Gnedin:2011kj,Baldauf:2011bh}.   It needs only to be calibrated
once {for a given cosmological model}
since the small scale response is the same regardless of the survey geometry.
In this paper, we perform this calibration and test the accuracy with which it describes the SSC effect directly with
survey-windowed simulations \cite{HuWhite:01,Takahashietal:09,dePutter:2011ah}.

The outline of this paper is as follows.   In \S \ref{sec:ssc}, we review the origin of
SSC, its description as the power spectrum response to a background mode, and uncover
a new dilation effect which changes scales in the power spectrum coherently within the
survey using the halo model.   We detail the separate universe calibration of the response
in \S \ref{sec:separate} which automatically captures all effects.   In \S \ref{sec:sscsim}, we 
test this calibration against direct super-survey sized simulations of the SSC effect.
We discuss these results in \S \ref{sec:discussion}.

\section{Super-Sample Covariance}
\label{sec:ssc}

Here we briefly review the theory for SSC that was
developed in Ref.~\cite{Takada:2013wfa}.   In \S \ref{sec:ssctri} we discuss its relationship
to squeezed trispectra and the separate universe consistency ansatz.  We examine in \S \ref{sec:sschalo} the halo model for SSC and introduce a dilation effect, the modulation of  
a short distance scale in a long wavelength mode, that is new to this work.

\subsection{Power spectrum covariance}
\label{sec:ssctri}

In a finite survey volume, we effectively measure 
 the underlying density fluctuation field $\delta(\bx)$
through a survey window function $W(\bx)$ which is 1 in the measured
region and 0 in the unmeasured region.  
The power spectrum, defined through
\begin{equation}
\langle \tilde \delta(\bk ) \tilde \delta(\bk') \rangle = (2\pi)^3 \delta_D^3(\bk+\bk') P(k),
\end{equation}
can be estimated 
as
\begin{equation}
\hat P(k_i) \equiv \frac{1}{V_W} \int_{|\bk|\in k_i}\!\! \frac{d^3
 \bk}{V_{k_i}} 
\tilde\delta_W(\bk)\tilde\delta_W(-\bk).
\label{eq:ps_est}
\end{equation}
Here  tildes represent the Fourier transform of a real space field,
the integral is over a shell in $k$-space of width $\Delta k$,
volume $V_{k_i}\approx 4\pi k_i^2 \Delta k$ for $\Delta k /k_i \ll 1$, and
 the effective survey volume is defined as
\begin{equation}
 V_W\equiv \int\!d^3\bx~W(\bx).
\end{equation}
The ensemble average of its estimator is a convolution of the underlying power spectrum 
with the window
\begin{eqnarray}
\!\!\!\!\!\!
\langle \hat P(k_i) \rangle &=& \frac{1}{V_W}   \int_{|\bk|\in 
k_i}\!\! \frac{d^3 \bk}{V_{k_i}}  \int\!\! \frac{d^3 \bq}{(2\pi)^3}  
\big|\tilde{W}(\bq)\big|^2 P(\bk-\bq).
\label{eqn:Pwindowed}
\end{eqnarray}
The survey window has support for $q \lesssim 1/L$ where $L = V_W^{1/3}$ is the typical
size of the survey.   Thus
for $k \sim 1/L$ this estimator is biased low compared to the true power spectrum
due to transfer of power into the fluctuation in the spatially-averaged density of
the survey volume {\citep{Takahashietal:09}}.    For $k \gg 1/L$ this
bias becomes progressively smaller since the underlying power spectrum is expected to 
be smooth across $\Delta k \sim 1/L$.   In \S \ref{sec:SSCpower}, we verify these properties with power spectra estimation in
simulations.

Due to the convolution, the window also increases the covariance of power spectrum estimators
separated by $\Delta k \sim 1/L$ and decreases their variance. On the other
hand, we are interested in 
the covariance of the power spectrum estimator which is induced  
across scales $\Delta k \gg 1/L$ dynamically via mode coupling in the density evolution.   In this wide bin limit, the covariance becomes
\begin{eqnarray}
C_{ij}&\equiv& 
\langle {\hat{P}(k_i)\hat{P}(k_j)} \rangle-
\langle {\hat{P}(k_i)}\rangle \langle{\hat{P}(k_j)}\rangle \nonumber\\
& \approx & C_{ij}^\text{G} + \frac{1}{V_W} \bar{T}_W(k_i,k_j).
\label{eq:pscov1}
\end{eqnarray}
Here the disconnected or Gaussian piece is
\begin{equation}
C_{ij}^\text{G} \equiv \frac{1}{V_W} \frac{(2\pi)^3}{V_{k_i}} 2 P(k_i)^2
			     \delta^K_{ij},
			     \label{eqn:Cgaussian}
\end{equation} 			     
with $\delta_{ij}^K=1$
if $k_i=k_j$ to within the bin width, otherwise $\delta_{ij}^K=0$.
 The
second term, proportional to $\bar{T}_W(k_i,k_j)$, is the non-Gaussian
contribution arising from the connected 4-point function or trispectrum
\begin{eqnarray}
\langle \tilde \delta(\bk_1 ) \tilde \delta(\bk_2 )\tilde \delta(\bk_3
 )\tilde \delta(\bk_4 )\rangle_c &&\nonumber\\
&&\hspace{-5em}=
(2\pi)^3 \delta_D^3(\bk_{1234}) T(\bk_1,\bk_2,\bk_3,\bk_4),
\end{eqnarray}
convolved with the survey
window function:
\begin{eqnarray}
 \bar{T}_W(k_i,k_j)&=&\frac{1}{V_{W}}
\int_{|\bk|\in k_i}\!\!\frac{d^3\bk}{V_{k_i}}
\int_{|\bk|'\in k_j}\!\!\frac{d^3\bk'}{V_{k_j}}\nonumber\\
&&\hspace{-3em}\times\int\!\left[\prod_{a=1}^4\frac{d^3\bq_a}{(2\pi)^3}
\tW(\bq_a)\right](2\pi)^3\delta^3_D(\bq_{1234})\nonumber\\
&&\hspace{-3em}\times
T(\bk+\bq_1,-\bk+\bq_2,\bk'+\bq_3,-\bk'+\bq_4).
\label{eq:pscov}
\end{eqnarray}
Here and below the notation $\bq_{1\ldots n}= \bq_1+\ldots+ \bq_n$.
The convolution with the window function means that 
 different 4-point configurations separated by less than the Fourier
width of the window function contribute to the covariance.  We call this aspect of the covariance the SSC effect.

\subsection{Trispectrum Consistency}

The trispectrum consistency condition introduced in Ref.~\cite{Takada:2013wfa} asserts
that the SSC term in the trispectrum must be consistent with the response of the
power spectrum to change in the background density by a factor of
$(1+\delta_b)$:
\begin{eqnarray}
\bar T(\bk,-\bk+\bq_{12},\bk',-\bk'-\bq_{12})
&\approx&  {T}(\bk,-\bk,\bk',-\bk')\nonumber\\
&&\hspace{-5em}+
\frac{\partial P(k)}{\partial \delta_b} \frac{\partial P(k')}{\partial \delta_b} P_\lin(q_{12}),
\label{eq:consistency}
\end{eqnarray}
where we have assumed $k, k'\gg q_{12}$.  The overbar here refers to an
angle average over the direction of $\bq_{12}$. 
Here $P_\lin$ is the
linear power spectrum and is designated as such to remind the reader
that for this relation to be applicable $\delta_b$ must be a mode in the
linear regime, i.e.\ the survey scale must be much larger than the nonlinear scale.
With this consistency prescription, the power spectrum covariance is given by %
\begin{equation}
C_{ij} = C_{ij}^{\rm G} + C_{ij}^{\text{T}0} 
+\sigma_b^2 \frac{\partial P(k_i)}{\partial \delta_b} \frac{\partial P(k_j)}{\partial \delta_b},
\label{eq:covcon}
\end{equation}
where we have introduced the variance of the background density
field $\delta_b$ in the survey window, defined as
\begin{equation}
\sigma_b^2 \equiv \frac{1}{V_W^2}
\int\!\!\frac{d^3\bq}{(2\pi)^3}
|\tW(\bq)|^2P_\lin(q).
\label{eq:sigmw}
\end{equation}
Here
\begin{equation}
 C^{\text{T}0}_{ij}\equiv\frac{1}{V_W}
\int_{|\bk|\in k_i}\!\!\frac{d^3\bk}{V_{k_i}}
\int_{|\bk'|\in k_j}\!\!\frac{d^3\bk'}{V_{k_j}}T(\bk,-\bk,\bk',-\bk'),
\end{equation}
is the non-Gaussian covariance term induced by nonlinearity that
is not mediated by a long wavelength mode 
 \citep{Scoccimarroetal:99}.
The linear variance $\sigma_b$ can be easily computed
for any survey geometry, either by evaluating Eq.~(\ref{eq:sigmw}) directly, or using
Gaussian realizations of the linear density field.

Note that we define density fluctuations relative to the global mean density rather
than the mean within the survey window  \cite{dePutter:2011ah}
\begin{equation}
P_W(k) =  P(k)/(1+\delta_b)^2.
\label{eqn:PW}
\end{equation}
$P(k)$ is appropriate for statistics such as weak lensing,
 where only the matter 
density is involved and its background value is fixed by cosmological
parameters, whereas
$P_W(k)$ characterizes statistics such as galaxy clustering where the mean density of
tracers is determined from the survey itself.
For the covariance
of $P_W(k)$, one would simply rescale the response as 
\begin{equation}
\frac{\partial P(k)}{\partial \delta_b} \rightarrow
\frac{\partial  P_W(k)}{\partial \delta_b} \approx
 \frac{\partial  P(k)}{\partial \delta_b}- 2 P(k).
 \label{eqn:localresponse}
\end{equation}
We numerically calibrate the response of both $P(k)$ and $P_W(k)$ in this way with separate universe simulations in 
\S \ref{sec:separate}.

\subsection{Halo Model}
\label{sec:sschalo}

For comparison to simulations and a physical understanding of results,  it is useful to have a semi-analytic model for SSC
and the response of the power spectrum to a background mode.    We review the halo
model construction of the trispectrum here and identify a new effect that was missing
in previous treatments \cite{Hamiltonetal:06,Takada:2013wfa}.

In the halo model
 \cite{PeacockSmith:00,Seljak:00,MaFry:00,CooraySheth:02}, 
 the power spectrum itself is described as
\begin{equation}
P(k) = P_{1h}(k) + P_{2h}(k),
\end{equation}
where the first term involves two points correlated by being in the
 same halo 
 \begin{equation}
 P_{1h}(k)=I_2^0(k,k)
 \end{equation}
 and the second term, two points in separate halos that are themselves correlated
 by the linear power spectrum
 \begin{equation}
P_{2h}(k)= [I_1^1(k)]^2 P_\lin(k).
 \end{equation}
We use the general notation \cite{CoorayHu:01}
 \begin{equation}
 I^\beta_\mu(k_1,\dots,k_\mu)\equiv 
\int\!\!dM\frac{dn}{dM}\left(\frac{M}{\bar{\rho}_m}\right)^\mu b_\beta  \prod_{i=1}^\mu
 \tu_M(k_i),
\end{equation}
where $M$ is the halo mass, $dn/dM$ is the halo mass function, $\bar\rho_m$ is the background
matter density, $b_0=1$,  $b_1=b(M)$ is the linear halo bias, and $\tilde u_M(k)$ is the Fourier transform of the halo density profile
normalized so that $\tilde u_M(0)=1$.

Specifically for calculational purposes we employ a Navarro-Frenk-White halo profile
\citep{Navarroetal:97} with the concentration-mass relation   \cite{Duffyetal:08},
\begin{equation}
c(M)=7.43(1+z)^{-0.71}\left(\frac{M}{M_\ast}\right)^{-0.081}.
\end{equation}
Here 
 the nonlinear halo mass scale today $M_\ast=3.91\times
10^{12}~M_\odot/h$ for our fiducial cosmological model (see Tab.~\ref{tab:LCDMpar}). For the halo
mass function and the halo bias, we employ the fitting formula in
Ref.~\cite{Bhattacharyaetal:10}. Note $b(M)\simgt 1$ when $M>M_\ast$ and this translates into a physical scale through the concentration-mass relation.

The halo model describes the matter trispectrum in a similar fashion but
with contributions involving one to four halos:
\begin{equation}
T=T_{1h}+\left(T_{2h}^{22}+T_{2h}^{13}\right)+T_{3h}+T_{4h}, 
\end{equation}
where $T_{nh}$ denote the $n$-halo terms whereas the superscripts on $T_{2h}$ denote the number of points in each of the 2 halos. These are
given by \cite{CoorayHu:01,Takada:2013wfa}
\begin{eqnarray}
 T_{1h}&=&
I^0_4(k_1,k_2,k_3,k_4),\\
T_{2h}^{22} &=&
P_\lin(k_{12})I^1_{2}(k_1,k_2)I^1_{2}(k_3,k_4)+
\mbox{2 perm.},\nonumber\\
T_{2h}^{13} &=&
P_\lin(k_1)I_1^1(k_1)I^1_{3}(k_2,k_3,k_4)+\mbox{3 perm.},\nonumber\\
 T_{3h}&=&
B_{\rm PT}(\bk_1,\bk_2,\bk_{34})
I^1_1(k_1)I^1_1(k_2)I^1_2(k_3,k_4)
+\mbox{5 perm.},\nonumber\\
T_{4h}&=& T_{\rm PT}(\bk_1,\bk_2,\bk_3,\bk_4)
I^1_1(k_1)I^1_1(k_2)I^1_1(k_3)I^1_1(k_4), \nonumber
\end{eqnarray}
where we have omitted the $\bk_i$ arguments of $T$ for brevity.   
$B_{\rm PT}$ and $T_{\rm PT}$ are the bispectrum and trispectrum
predicted from the perturbation theory \cite{Bernardeauetal:02}. 
{Note that the 1 halo term is strongly weighted to the highest masses or rarest halos
in the mass function where the ingredients of the halo model are the least well calibrated.}

For squeezed configurations with $k,k'\gg q_{12}$, we can express the change in the
trispectrum due to the long wavelength $q_{12}$ mode to leading order in $q_{12}/k$ 
as
\begin{equation}
\delta T \equiv T(\bk,-\bk+\bq_{12},\bk',-\bk'-\bq_{12})-T(\bk,-\bk,\bk',-\bk')
\end{equation}
with
\begin{eqnarray}
\delta T_{1h}
&\approx & 0,
\nonumber\\
 \delta T_{2h}^{22} &\approx & 
P_\lin(q_{12})I^1_2(k,k)I^1_2(k',k'),
\nonumber\\
 \delta T_{2h}^{13} &\approx &  0,
\nonumber\\
 \delta  T_{3h} 
&\approx& 
2P_\lin(q_{12}) I^1_2(k',k') \fpole(\bk,\bq_{12}) \nonumber\\
&& + 
2P_\lin(q_{12}) I^1_2(k,k) \fpole(\bk',-\bq_{12}),
\nonumber\\
\delta T_{4h}
&\approx&  
4 P_\lin(q_{12})  \fpole(\bk,\bq_{12})  \fpole(\bk',-\bq_{12}),
\label{eq:Thm}
\end{eqnarray}
where
\begin{eqnarray}
\fpole(\bk,\bq) &\equiv&
 [ P_\lin(k) F_2(\bq,-\bk) + P_\lin(|\bk -\bq|) F_2(\bq,\bk-\bq) ] \nonumber\\
&& \times  I_1^1(k) I_1^1(|\bk -\bq|),
\end{eqnarray}
with the 
mode-coupling kernel $F_2$ is defined as
\begin{align}
F_2(\bk,\bq) \equiv &\frac{5}{7}+\frac{1}{2}
\left(\frac{1}{k^2}+\frac{1}{q^2}\right)(\bk \cdot\bq)
+\frac{2}{7}\frac{(\bk \cdot\bq)^2}{k^2 q^2}.
\end{align}

\wh{
This expression differs from {Eq.~(32)}  of Ref.~\cite{Takada:2013wfa} 
in that we have 
made approximations such as $\bk -\bq_{12} \approx \bk$ only in terms that do not
involve  $\fpole$.}  
The latter must be handled with care since 
the mode coupling factor $F_2$ has a pole when one of its arguments
goes to zero.    Thus we need to consistently expand this expression in $q_{12}/k$ (or 
$q_{12}/k'$).   
The result of integrating over the direction of $\bk$ is 
 \begin{eqnarray}
\int^{1}_{-1}\!\frac{d\mu_k}{2}
\fpole (\bk,\bq) &\approx & \frac{1}{2}\left(
\frac{68}{21}-\frac{1}{3}\frac{d\ln k^3 P_{2h}}{d\ln k}
\right)P_{2h}.
\end{eqnarray}
It is now straightforward to see that with the association 
\begin{equation}
\frac{\partial  P(k)}{\partial  \delta_b} 
\approx {\left(\frac{68}{21}-\frac{1}{3}\frac{d\ln k^3P_{2h}(k)}{d\ln k}\right)
P_{2h}(k)
+ I_2^1(k,k)},
\label{eq:haloresponse}
\end{equation}
the terms involving $P_\lin(k)$ take the SSC form.
The $68/21$ piece is called
the beat coupling (``BC'') effect that the growth of a short wavelength perturbation is enhanced in a large scale overdensity, the $I_2^1$ term is the halo sample variance (``HSV'') effect
that halo number densities are also enhanced in such a region,
and the derivative term is new to this work which we call the linear dilation (``LD'') effect.
The dilation effect occurs because the long wavelength mode changes the local expansion factor and hence the physical size of a mode that would have comoving wavelength $k$ in its absence (see \S \ref{sec:sepder}).

\begin{figure}[tb]
\centering
    \includegraphics[width=3.4in]{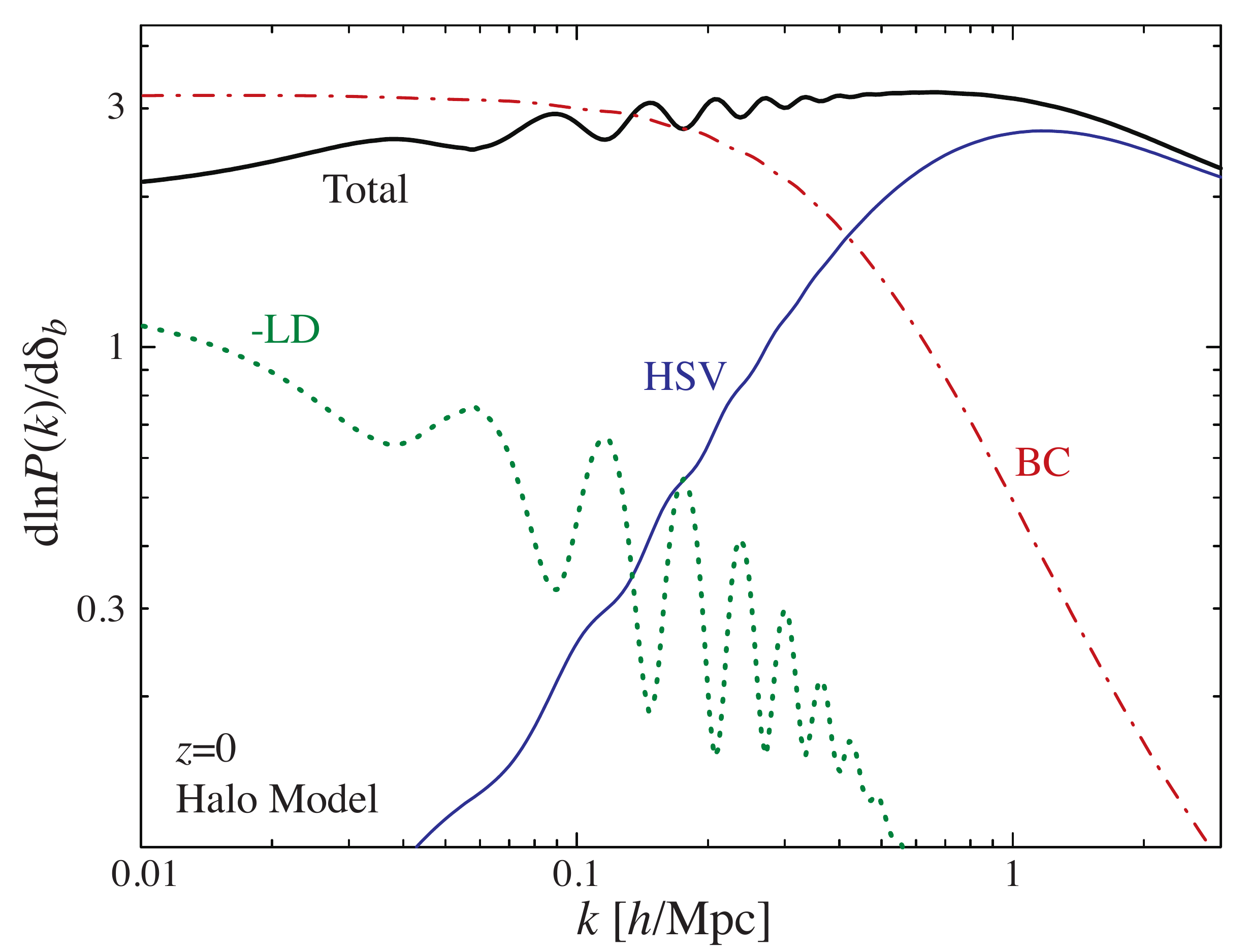}
    \caption{\footnotesize Power spectrum response to a long-wavelength background mode in the halo
 model separated into three contributions {(see Eq.~\ref{eq:haloresponse})}:  beat coupling (BC, change in short wavelength linear growth),  halo sample
 variance (HSV, change in the number density of halos), and the linear
 dilation (LD,  change in scale in the linear regime). The LD term is negative  and partially cancels the BC effect while also changing the scale of the baryon acoustic oscillations.  Dilation is included in HSV since the mass
    function is compatible with the Eulerian bias.  Cosmological parameters are set
    here according to Tab.~\ref{tab:LCDMpar}.
    }
    \label{fig:halo_terms}
\end{figure}

In Fig.~\ref{fig:halo_terms} we show this halo model decomposition of the response.  
Note that in the linear regime the LD term cancels part of the BC contribution and
lowers the overall effect.   The LD term also responds to the baryon acoustic oscillations (BAO)
and represents the fact that dilation changes the scale of any features in the power spectrum.  
In the presence of SSC, the BAO scale will vary fractionally by  $\sim \sigma_b$ from 
volume to volume or bias results from a single volume by that amount
(see also \cite{Crocce:2007dt,SherwinZaldarriaga:12} for the residual impact 
 after averaging over volumes).

  In fact dilation affects the 1-halo HSV term as
well: in the peak-background split, the number density of halos is enhanced by two factors \cite{Moetal:97}.
The first is that the threshold for collapse is lowered in the presence of the long wavelength
mode leading to bias in Lagrangian space.    The second is the change in physical
scales, common to both the halos and matter density,
due to the collapse of the large scale region that converts Lagrangian bias to Eulerian bias.   The latter is the dilation effect and hence it is automatically included for halos by the fact that the Eulerian 
bias $b$ is consistent with the mass function
\begin{equation}
\int\!\!dM\frac{dn}{dM} \frac{M}{\bar{\rho}_m} b(M) = 1,
\end{equation}
which enforces
\begin{equation}
\frac{ \partial I_2^0}{\partial \delta_b} = I_2^1
\label{eq:Iconsistency}
\end{equation}
so that the HSV term represents the total response to the background mode.

\section{Separate Universe Response}
\label{sec:separate}

In this section, we use the separate universe technique to calibrate numerically the 
response of the power spectrum to a background mode into the nonlinear regime.
In \S \ref{sec:sepparam}, we review the Newtonian simulation prescription following Ref.~\cite{Sirko:05} (see  \cite{Baldauf:2011bh,Gnedin:2011kj} for horizon scale generalizations
and \cite{1996ApJ...472...14T,1997MNRAS.286...38C,Schneider:2011wf} for related 
methods for restoring linear modes).
We describe two techniques for
differencing of power spectra to calibrate the background response in  \S \ref{sec:sepder} and
simulation details in  \S \ref{sec:sepsim}.   In \S \ref{sec:sepres} we present results for the power spectrum response in the separate universe approach.

\subsection{Separate Universe Parameters}
\label{sec:sepparam}
In the separate universe technique, the mean density fluctuation in the survey window
$\delta_b$ is absorbed into the background density
{$\bar{\rho}_{mW}$} 
of a separate  universe
\begin{eqnarray}
\bar{\rho}_m (1+\delta_b) &=&  \bar{\rho}_{mW}
\end{eqnarray}
so that the separate universe parameters obey
\begin{eqnarray}
\frac{\Omega_m h^2}{a^3}(1+\delta_b) &=& \frac{\Omega_{mW} h_W^{2}}{a_W^{3}},
\label{eqn:equatedensity}
\end{eqnarray}
where $H_0 = 100h$ km s$^{-1}$ Mpc$^{-1}$.  
Our convention is to set the scale factor of the separate universe $a_W$ to
agree with the global one at high redshift
\begin{equation}
\lim_{a\rightarrow 0} a_W(a,\delta_b)=a,
\end{equation}
since 
\begin{equation}
\lim_{a\rightarrow 0} \delta_b(a)=0.
\end{equation}
Thus the description and physical content of the separate universe is the same at high redshift
\wh{and so
\begin{eqnarray}
\Omega_{mW} h_W^2 &=& \Omega_m h^2.
\label{eqn:omegamh2}
\end{eqnarray}
In fact  this equality says that 
the {\it background} densities of the two universes $\bar\rho_m$ and
$\bar\rho_{mW}$ are always the same 
 at the same numerical value of $a$ and $a_W$.
Conversely, by virtue of Eq.~(\ref{eqn:omegamh2}), an equal time (or equal {\it survey} density) comparison as in Eq.~(\ref{eqn:equatedensity}) corresponds to different scale factors
\begin{eqnarray}
a_W  = \frac{a}{(1+\delta_b)^{1/3}}&\approx& a\left(1- \frac{\delta_b}{3} \right)
\label{eqn:aW}
\end{eqnarray}
in the respective cosmologies.}
The separate universe also has a different expansion rate since the peculiar velocities implied
by continuity from the linear density evolution are reabsorbed into the background expansion.
Using the definition $H = \dot a/a$ and Eq.~(\ref{eqn:aW}), we obtain
\begin{equation}
\delta H^2 = H_W^2 - H^2 \approx -{\frac{2}{3} }H \dot \delta_b .
\label{eqn:dh2}
\end{equation}
In a $\Lambda$CDM universe, the growth rate is given by
\begin{equation}
H \dot \delta_b = \frac{\Omega_m H_0^2}{2 a^2} \left[ \frac{5}{D} -\frac{3}{a} - \frac{2 \Omega_K}{\Omega_m} \right] \delta_b,
\label{eqn:ddeltab}
\end{equation}
where the linear growth function $\delta_b = (D/D_0)\delta_{b0}$ is
normalized to 
\begin{equation}
\lim_{a\rightarrow 0} D = a.
\end{equation}

We can then match this perturbation to the separate universe Friedmann equation
\begin{eqnarray}
H_W^2 &=& H_{0W}^2 \left[ \frac{\Omega_{mW}}{a_W^3} + \Omega_{\Lambda W} 
+ \frac{\Omega_{KW}}{a_W^2} \right] \\
&\approx& H^2 + \frac{H_{0W}^2-H_0^2}{a^{2}} +H_0^2\delta_b\left[
\frac{\Omega_m}{a^{3}}+\frac{2}{3}\frac{\Omega_K}{a^{2}}\right] \nonumber
\end{eqnarray}
 to
define its Hubble constant 
\begin{equation}
H_{0W}\equiv H_W\big|_{a_W=1} \ne H_W \big|_{a=1},
\end{equation}
where we have used  Eqs.~(\ref{eqn:omegamh2}), (\ref{eqn:aW}), $\sum_i \Omega_{iW}=1$ and the fact that
 the cosmological constant is a constant physical density.
Comparing this equation with Eq.~(\ref{eqn:dh2}), we obtain  the separate universe  Hubble
constant as \cite{Sirko:05}
\begin{equation}
 \frac{\delta h}{h}  \equiv \frac{ H_{0W} - H_0}{H_0} \approx -\frac{5\Omega_m}{6}\frac{\delta_b}{D}.
 \label{eqn:hpert}
\end{equation}
Since $\delta_b/D= {\delta_{b0}/D_0}$, this relation holds independently of the redshift at which $\delta_b$ and $D$ are
defined.  

The other parameters then directly follow from this relation and
Eq.~(\ref{eqn:omegamh2})
\begin{eqnarray}
\frac{\delta \Omega_m}{\Omega_m}
=\frac{\delta \Omega_\Lambda}{\Omega_\Lambda}
 &\approx& 
{- 2 \frac{\delta h}{h}.}
\end{eqnarray}
Even if the global universe is flat, the separate universe would have
spatial curvature, since
\begin{eqnarray}
\Omega_{KW}&=& 1-\Omega_{mW}-\Omega_{\Lambda W} \nonumber\\
&=& 1-(\Omega_{m}+\Omega_{\Lambda})\left(1+\frac{5\Omega_m}{3}\frac{\delta_b}{D}\right),
\end{eqnarray}
yielding a closed separate universe for $\delta_b>0$.  

\subsection{Power Spectrum Derivatives}
\label{sec:sepder}

With the separate universe cosmological parameters set, we can
conduct simulations to calibrate the response of the power spectrum by finite difference
of models with $\delta_b =\pm \epsilon$ where $\epsilon \ll  1$.   We discuss here
a number of further subtleties as to how this comparison is performed 
namely the choice of what is held fixed when differencing.

 First let us employ the convention in this
section that $k$-values are quoted in comoving Mpc of their respective cosmologies.   $N$-body codes are typically written with $k$ in $h\,$Mpc$^{-1}$ and so it is to be understood here that
in those units the box scale and modes carry an $h/h_W$ conversion factor to fix scales
in Mpc (see \S \ref{sec:sepsim}).  To reduce confusion, we work with
$\Delta_W^2 = k_W^3 P_W(k_W)/2\pi^2$ which is dimensionless and independent of units.

Although both $k$ and $k_W$ are in comoving Mpc$^{-1}$ in their respective cosmologies, we compare the scales at different scale factors given by Eq.~(\ref{eqn:aW}).   This means that the physical scale of the separate universe  $k_W$ corresponds to the global  $k$ as
\begin{equation}
k_W(k,\delta_b) = \frac{a_W}{a} k \approx \left( 1- \frac{\delta_b}{3} \right) k.
\label{eqn:kW}
\end{equation}
Furthermore the power spectrum with fluctuations referenced to the global mean is
related to that of the local power spectrum by
\begin{equation}
\Delta^2(k) = (1+\delta_b)^2 \Delta_W^2(k_W,\delta_b)
\end{equation}
corresponding to the rescaling in Eq.~(\ref{eqn:PW}).  \wh{Again using dimensionless power spectra $\Delta_W^2$ avoids the potential confusion that the numerical value of the power 
spectrum $P_W$ depends on the units with which it is measured so that
$k^3 P_W(k) \equiv k_W^3 P_W(k_W)$.}
We now evaluate the power spectrum response to $\delta_b$ as
\begin{eqnarray}
\frac{d\ln P}{d\delta_b} = \frac{d \ln \Delta^2}{d\delta_b} 
= 2 + \frac {d \ln \Delta_W^2(k_W,\delta_b)}{d\delta_b}\Big|_k.
\label{eqn:totalderiv}
\end{eqnarray}
Thus we seek to evaluate the change with $\delta_b$
 in the separate universe power spectrum at fixed $k$.

There are multiple ways in which we can evaluate this change by finite differencing
the power spectra of separate universe simulations.    In a simulation, the Gaussian random
initial conditions introduce stochasticity in the power spectrum.  To reduce this, we wish
to difference modes in the simulations with the same initial seeds.  

In the separate universe construction, if we start the
two simulations with the same comoving box size then due to Eq.~(\ref{eqn:kW}) for the
 relationship 
between $k_W$ and $k$, we difference modes at different $k_W$.   To fix this problem
we can set the comoving box size to have the same physical scale at the epoch that
we want the derivative.   Then choosing the same seeds in code coordinates guarantees
that we are differencing modes with the same Gaussian random realization.
We call this  the total derivative technique.

One drawback of this technique is that each evaluation epoch requires a new set of
separate universe simulations even though fundamentally all epochs share the same separate
universe construction as noted below Eq.~(\ref{eqn:hpert}).   The alternative is to chain rule the derivative
\begin{eqnarray}
\label{eqn:nodilation}
&&\frac {d \ln \Delta_W^2(k_W,\delta_b)}{d\delta_b}\Big|_k\\
 &&\qquad= \frac {\partial \ln \Delta_W^2(k_W,\delta_b)}{\partial \delta_b}\Big|_{k_W} +  \frac {\partial \ln \Delta_W^2(k_W,\delta_b)}{\partial \ln k_W} \frac{\partial \ln k_W}{\partial\delta_b} 
\nonumber\\
&&\qquad \approx  \frac {\partial \ln \Delta_W^2(k_W,\delta_b)}{\partial \delta_b}\Big|_{k_W} -\frac{1}{3}  \frac {\partial \ln \Delta_W^2(k_W,\delta_b)}{\partial \ln k_W} \nonumber\\
&&\qquad \approx\frac {\partial \ln \Delta_W^2(k_W,\delta_b)}{\partial \delta_b}\Big|_{k_W} -\frac{1}{3}  \frac {d \ln \Delta^2}{d \ln k}. \nonumber
\end{eqnarray}
Now the differences can be taken at fixed separate universe comoving scales $k_W$ which means that the Gaussian random realizations can also be taken to be exactly the same at
high redshift.
Note also that in the linear regime this split is the division between  beat coupling
and dilation.  The first term represents the enhancement of linear or nonlinear growth
while the second term represents the dilation of the volume due to the different local expansion factor in the separate universe.    We thus call this the growth-dilation
derivative technique.

  In the absence of realization-to-realization scatter in this derivative, the growth-dilation method would require in principle only two separate universe simulations to calibrate the response
 at all redshifts.   However we shall see in \S \ref{sec:sepres} that higher order effects make the response to $\delta_b$ depend somewhat on the realization of small scale power itself, especially 
 in the deeply non-linear regime.   
 To obtain the mean response then we average over many realizations of the sub-survey
 power.     
By adding the mean dilation term rather than the dilation term of the same realization
in Eq.~(\ref{eqn:nodilation}), we 
enhance the run-to-run scatter of the combination and require more simulations to determine the mean than in the total derivative approach of Eq.~(\ref{eqn:totalderiv}).   Which method is to be preferred depends on
how many redshifts one wishes to calibrate the response.

\begin{table}[t]
\centering
\begin{tabular}{| c | c |}
\hline
$\Omega_m$ & 0.286 \\
$\Omega_b$ & 0.047 \\
$h$&0.7 \\
$n_s$ & 0.96 \\
$\sigma_8$ & 0.82 \\
\hline
\end{tabular}
\caption{\footnotesize Flat $\Lambda$CDM model parameters used throughout.}
\label{tab:LCDMpar}
\end{table}
\subsection{Simulations}
\label{sec:sepsim}

Here we give further details on the simulation pipeline.    
Our baseline cosmology is the flat $\Lambda$CDM model with parameters
given in Tab.~\ref{tab:LCDMpar}.  From this baseline, we set the separate universe parameters according to \S \ref{sec:sepparam} with $\delta_b=\pm 0.01$.

We then determine the initial conditions for
the separate universe runs using CAMB \cite{Lewis:1999bs,Howlett:2012mh}.   Since the separate universes are identical
at high redshift with $k$ in Mpc$^{-1}$, we fix the initial power spectrum  $\Delta_{\cal R}^2= A_s (k/k_0)^{n_s-1}$ with $k_0=0.05\,$Mpc$^{-1}$ to be the same in each case.    The initial conditions generator 
2LPTIC \footnote{\href{http://cosmo.nyu.edu/roman/2LPT/}{http://cosmo.nyu.edu/roman/2LPT/}}  takes the $a_W=1$ power spectrum normalization $\sigma_{8W}$, the 
rms linear density in $8 h_W^{-1}$ Mpc radius spheres, and functional form to generate
initial conditions at  $a_{Wi}=0.02$ which are used by L-Gadget2  \cite{Springel:2005nw} to produce the
separate universe simulations.

\begin{figure}[t]
    \centering
    \includegraphics[width=3.5in]{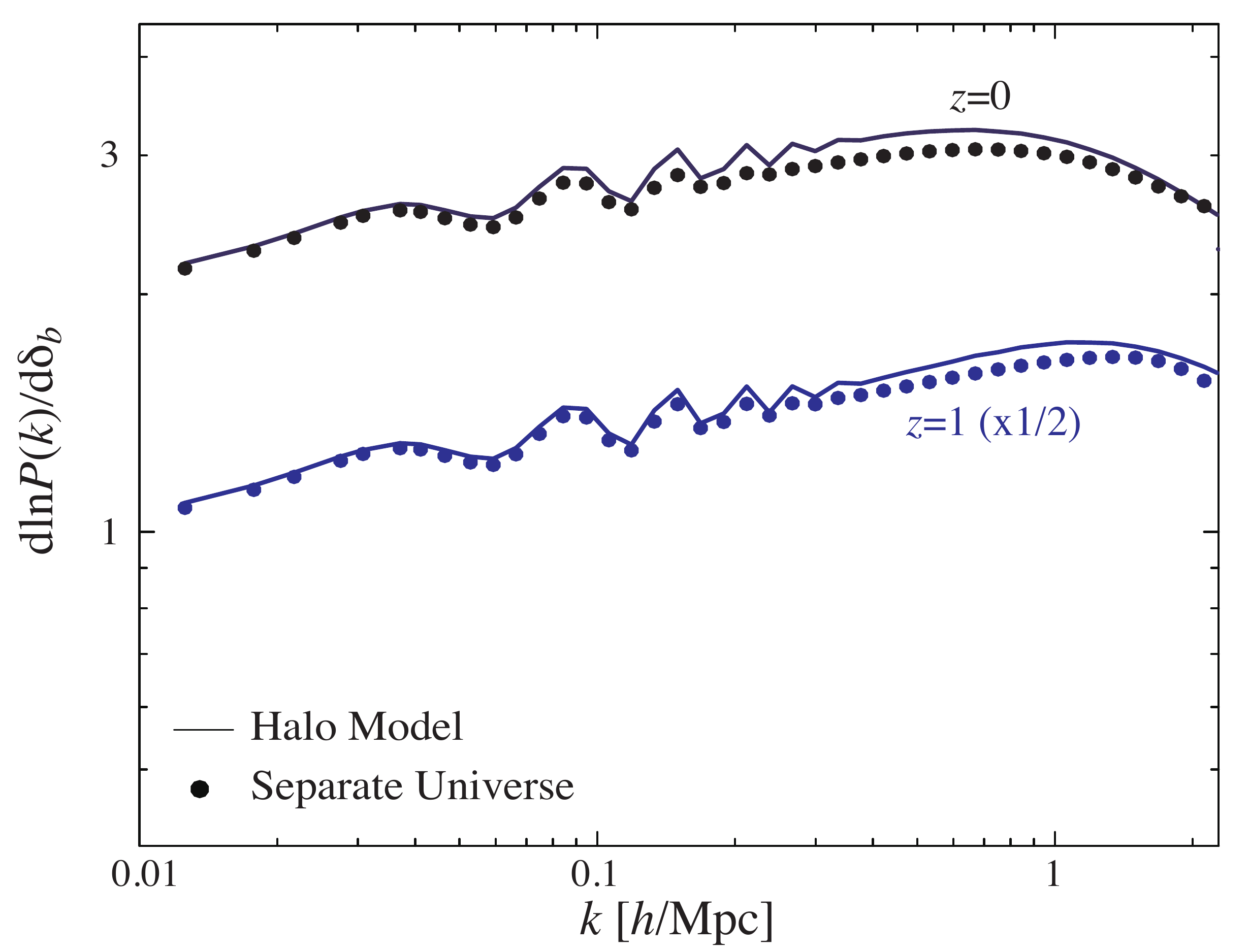}
    \caption{\footnotesize Separate universe (total derivative) technique vs halo model for the background response  at $z=0$ and $z=1$ (halved for clarity).  The halo model captures the qualitative features of the response.  Note that the halo model
    response differs from  that in Fig.~\ref{fig:halo_terms} since the power spectrum is binned with the same prescription as the separate universe results.}
    \label{fig:responsecomp}
\end{figure}

Both 2LPTIC and L-Gadget2 take the comoving scale of the simulation box $L_W$ to be in units of $h_W^{-1}$Mpc.   In the total derivative method, where we set the physical scale of the
simulations to coincide at the final evaluation epoch $a_{fW}$, we therefore set
\begin{eqnarray}
    a_{fW} \frac{L_{W}} {h_{W} }&=& a_f\frac{ L}{ h}, 
\end{eqnarray}
whereas for the growth-dilation method, where the comoving scale in Mpc is the same,
we set
\begin{equation}
    \frac{L_{W}} {h_{W}} = \frac{L}{ h},
\end{equation}
independently of the final epoch $a_f$ at which we want to calibrate the background
response.
In both cases we choose
\begin{equation}
    L = 500 \, [ h^{-1} {\rm Mpc} ].
\end{equation}

  In box coordinates we generate the same initial amplitudes and phases
for two runs with $\delta_b = \pm 0.01$.   We run L-Gadget2 with $256^3$ particles,
$512^3$  (Tree)-PM  grid and analyze the power spectrum by FFT with cloud in cell (CIC)
assignment to a $1920^3$ grid.   
{We bin the power spectrum estimates with $20$ logarithmically spaced bins per decade and show bins at the average $k$ weighted by the number of modes.}

 We then difference the binned $\Delta^2$ in box coordinates 
in each method to form the required derivative.   {To test the resolution dependence
of results, we have employed higher resolution simulations with $512^3$ particles and $1024^3$ (Tree)-PM grid to verify that at 
 $k \lesssim 2 h$/Mpc our response results have converged to several percent or better.}   Finally we simulate 64 separate universe pairs for each method for $a_{f}=1$
 to test the stochasticity of the
response.

\begin{figure}[tb]
    \centering
    \includegraphics[width=3.5in]{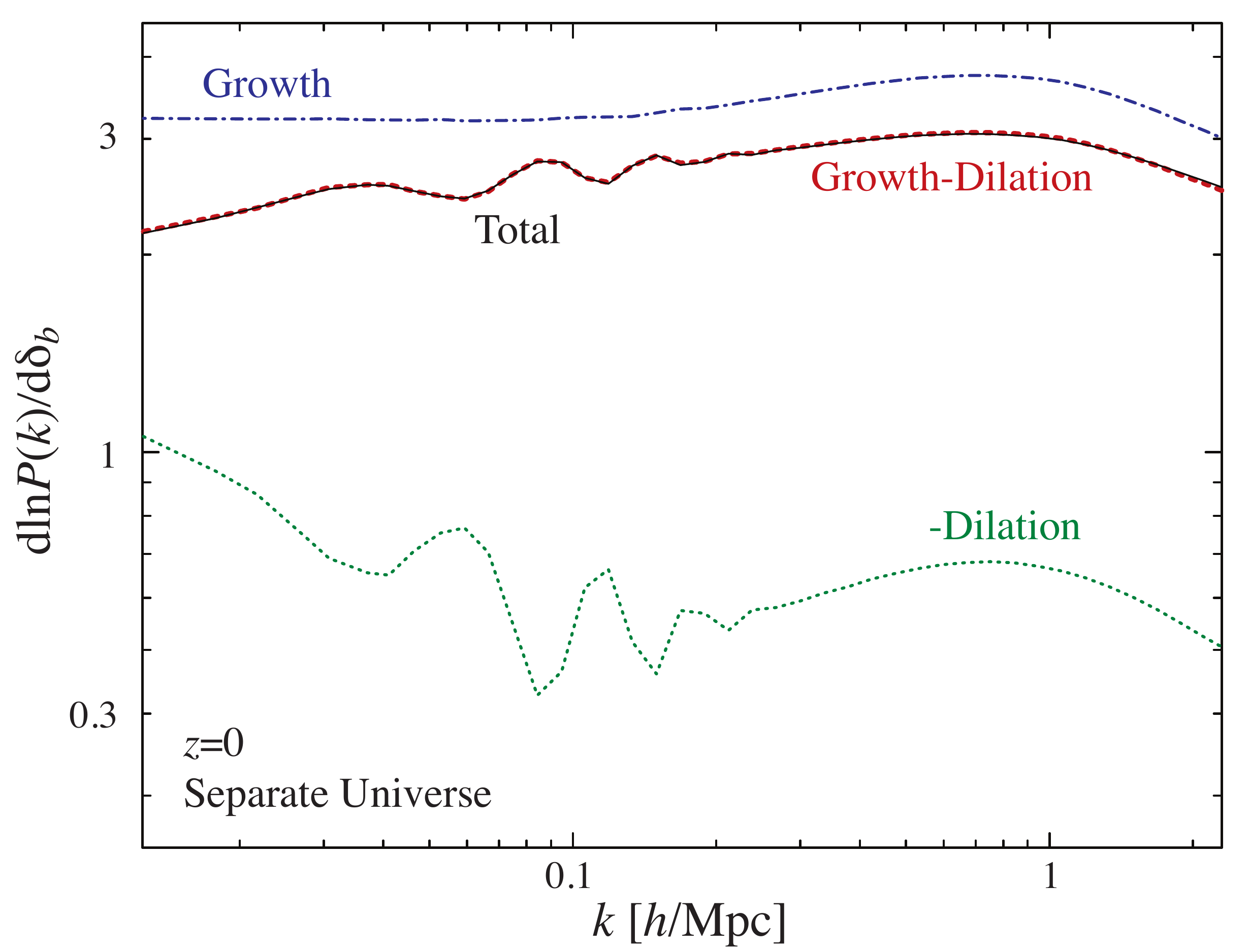}
    \caption{\footnotesize Separate universe total derivative and growth-dilation techniques for the power spectrum response to a background mode at $z=0$.   The two methods produce indistinguishable total derivatives once the growth and dilation (negative) pieces are combined.  Note that in the linear regime, the dilation contribution matches the LD term of
    the halo model in Fig.~\ref{fig:halo_terms} but differs in the nonlinear regime where it continues to cancel the growth term.}
    \label{fig:sim_terms}
\end{figure}

\subsection{Response Calibration}
\label{sec:sepres}

We choose as our primary technique the total derivative method in Eq.~(\ref{eqn:totalderiv}) averaged over  64 separate pairs as described in the previous section and test results
at $z=0,1$ against the halo model predictions (see Fig.~\ref{fig:responsecomp}).   Within
the domain of validity of the simulations, the halo model captures the
fractional response to the 6\% level or better.    
{Note that halo model predictions depend at this level of precision on 
the rarest halos which may not be well-represented in the simulations or well-calibrated
in the halo model employed.
}

The halo model shows a slightly larger  dilation term near the BAO scale at $z=0$
since there is no explicit mode coupling applied to the two halo term to smooth the intrinsic
features.   
Furthermore, for statistics such as galaxy clustering that are based on the power spectrum referenced to the survey
mean $P_W$ the response is lowered by an additive term $-2$ as in Eq.~(\ref{eqn:localresponse}) and hence the maximal difference is enhanced to  
16\%  in that case.   
Nonetheless the halo model serves to help physically interpret the
response.   For example the global response is allowed to fall below $2$ or local response below $0$ at high $k$ because
this region is dominated by halos whose  bias is less than unity.  The wavenumber at which this
occurs increases with redshift due to the increasing rarity of halos of fixed mass.

Using the growth-dilation technique, we can also further study the physical contributions to the response through separate universe simulations.   In Fig.~\ref{fig:sim_terms}, we show 
the growth and dilation contributions separately with this technique and demonstrate that 
the sum reproduces the total response.   Note that in the linear regime the dilation contribution
takes the same form as the halo model linear dilation term shown in Fig.~\ref{fig:halo_terms}
whereas in the nonlinear regime the dilation lowers the response from growth.   Recall that
in the halo model, dilation is automatically included in 
the halo sample variance term.

\begin{figure}[t]
    \centering
    \includegraphics[width=3.5in]{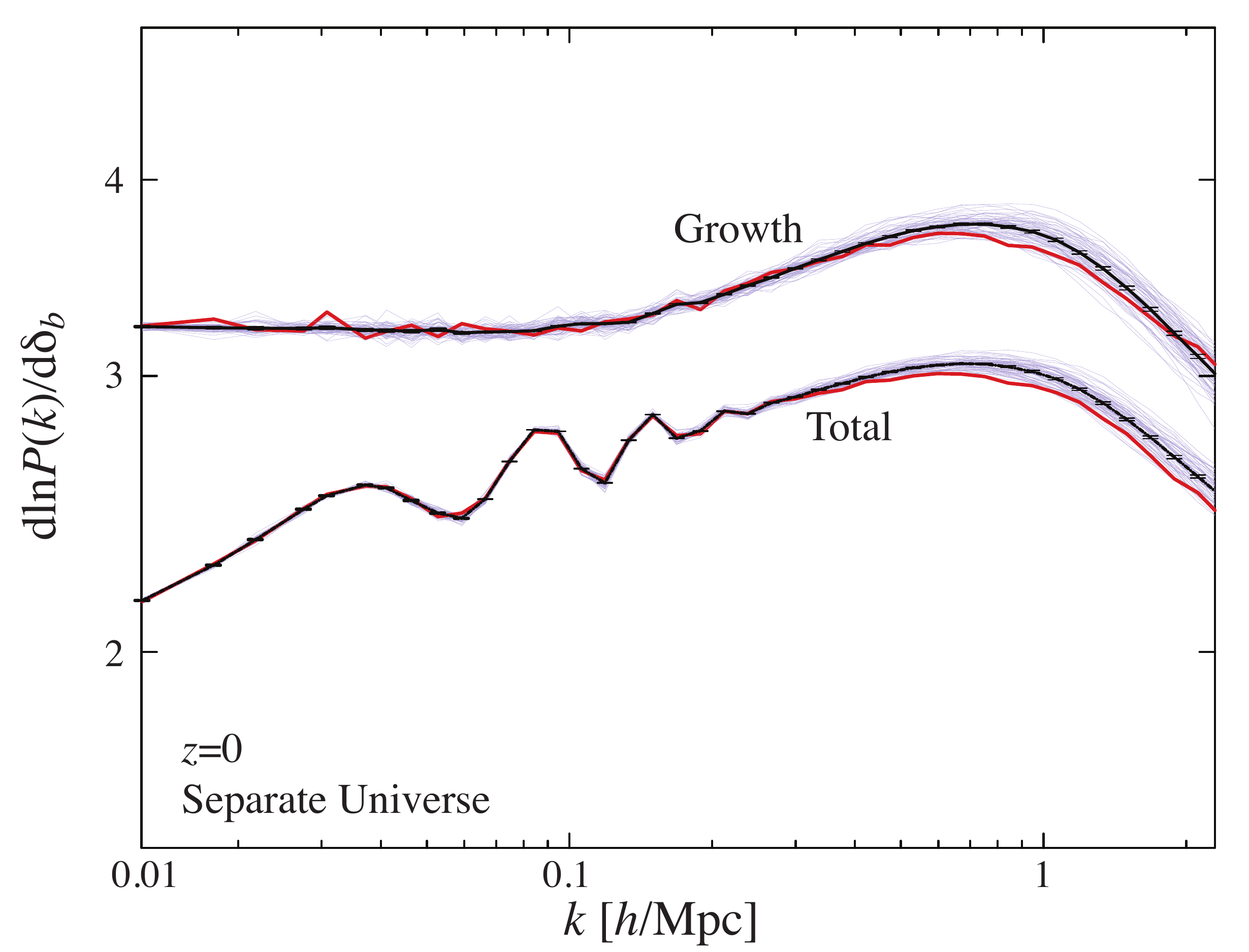}
    \caption{\footnotesize {Stochasticity in the separate universe background response for the finite difference of 64 pairs of simulations (thin curves) in the total derivative vs. growth only finite differences.    In the nonlinear region the fractional scatter (thin curves) is similar but the addition of the mean dilation term to form the total in the growth-dilation technique leads to larger scatter.  Means with standard errors and a highlighted single realization (thick curves) are shown here to illustrate that nonlinear deviations from the mean are highly correlated across $k$.}}
    \label{fig:dPdd_scatter}
\end{figure}

Finally in Fig.~\ref{fig:dPdd_scatter}, we show the stochasticity of the response  with
the 64 pairs of separate universe simulations in each technique.     Notice that the fractional
scatter in each technique is comparable in the nonlinear regime when analyzing the direct power spectrum
differences.   However in the growth-dilation technique, the mean dilation term cancels
the power spectrum difference leading to a larger fractional error on the total response.
Thus more pairs are required to obtain the total response to the same precision.  Conversely
the total technique requires separate pairs for each redshift.

In the absence of stochasticity, the SSC effect can be 
removed by fitting $\delta_b$ as a supplementary parameter to the usual cosmological
parameters $p_\text{cos}$ in a power spectrum model where
\begin{equation}
\langle \hat P(k;p_\text{cos},\delta_b)\rangle = P(k;p_\text{cos}) \left(1 + \frac{\partial \ln P}{\partial \delta_b} \delta_b \right),
\end{equation}
rather than including it as a covariance term \cite{Takada:2013wfa}.  While we have demonstrated that stochasticity is in practice present, it is sufficiently small that this model provides an excellent approximation in
the quasi-linear regime and remains a good approximation in the fully nonlinear regime.  Note that with the inclusion of the dilation term, this additional parameter is no longer
approximately degenerate with the power spectrum amplitude.

\section{SSC Simulation Tests}
\label{sec:sscsim}

In this section we compare the direct quantification of the SSC effect on the covariance
of the power spectrum in subvolumes of
a large volume simulation with the SSC model of the power spectrum response to
a background mode.  We begin in \S \ref{sec:SSCpower} with a discussion of the simulation suites, power
spectrum estimators, and bias correction for windowing effects.   We define 
power spectrum covariance estimators in \S \ref{sub:super_sample_covariance} and 
present results in \S \ref{sec:SSCresults}.

\subsection{Power Spectrum}
\label{sec:SSCpower}

We use a suite of 7 large-volume simulations originally made for the Dark Energy Survey.  Each corresponds to a $4h^{-1} \text{Gpc}$ box run with
 L-Gadget2, $2048^3$ particles,
$3072^3$   (Tree)-PM  grid and cosmological parameters of Tab.~\ref{tab:LCDMpar}.   We then CIC assign the particles to a  $(8 \times 1920)^3$ grid.
 We next subdivide each large box into $8^3=512$ subvolumes of size $500 h^{-1} \text{Mpc}$  each for a total of $N_{s}=3584$ subboxes.  
We  then extract the power spectrum of each of the subboxes by FFT before
deconvolving the CIC window and binning in $k$ to form $\hat P^\text{sub}$. 

{We choose a special binning scheme to minimize the effects of
convolution discussed in  \S \ref{sec:ssctri}.   These effects appear for $\Delta k
\,[h/\text{Mpc}] \sim
(2\pi/500 ) =0.0125$.    
For $k\,[h/\text{Mpc}] < 0.1$, we choose 5 approximately linearly spaced bins and
10 logarithmically spaced bins per decade above it.}

Finally we define the set of $N_s$ power spectra referenced to the subbox mean as
\begin{equation}
\hat P_W(k) = \frac{\hat P^\text{sub}(k)}{(1+ \delta_b)^2},
\end{equation}
where $\delta_b$ is the average density fluctuation in the same box.  Note that
at $z=0$, $\sigma_b=0.0126$ for these subboxes.

For comparison, we run a set $N_s$ small box simulations of size $500 h^{-1} \text{Mpc}$,
$256^3$ particles, and  $512^3$ (Tree)-PM  grid analyzed for $\hat P^\text{sm}$ with CIC assignment to a
$1920^3$ grid.   These settings match the $1/8$ scaling of the large box dimensions except
for the (Tree)-PM grid.

\begin{figure}[t]
    \centering
    \includegraphics[width=3.5in]{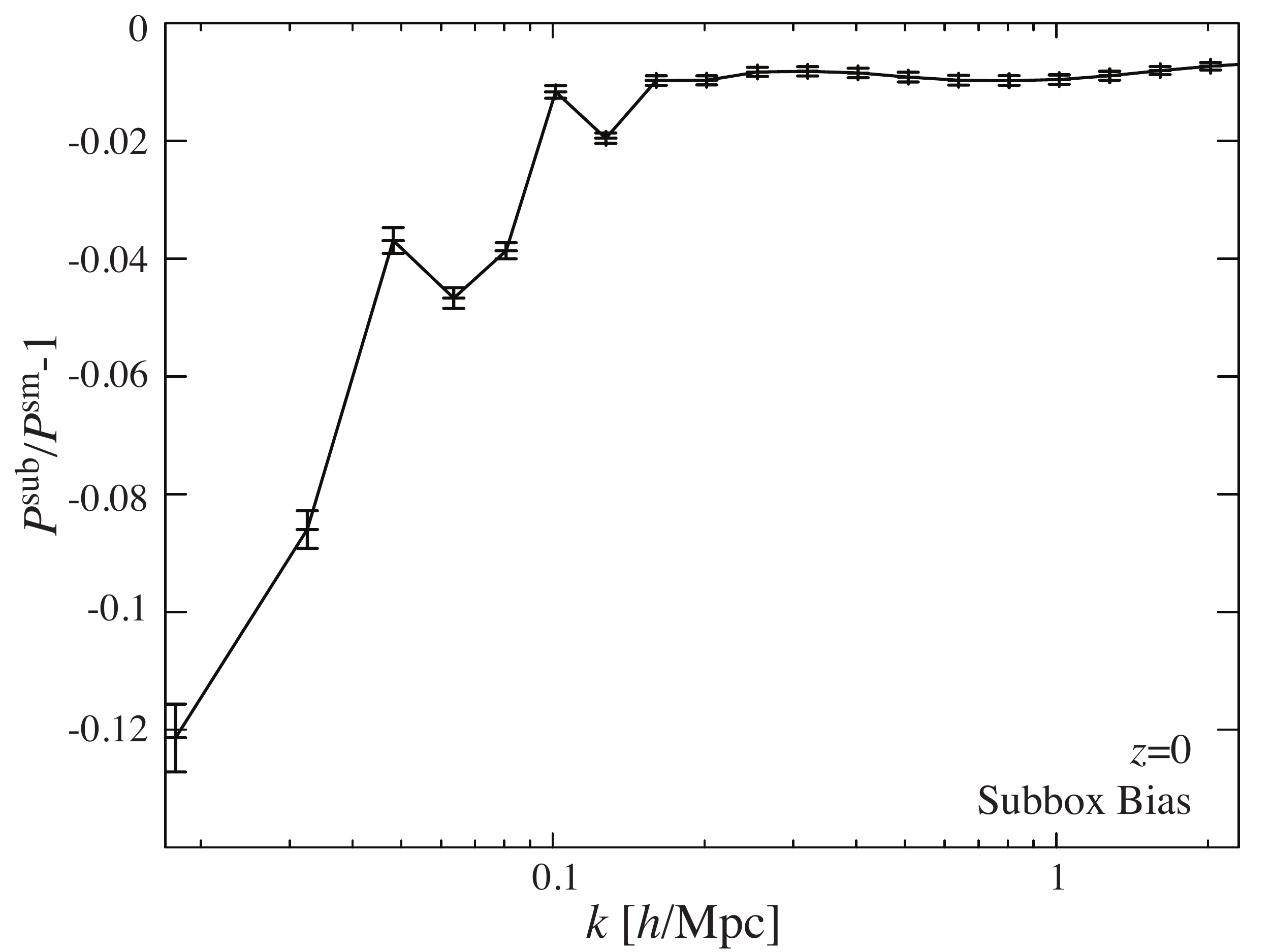}
    \caption{\footnotesize Bias between the subbox vs.\ small box  power spectrum estimators.
    The subbox power spectrum is a convolution of the true power spectrum with the survey window.  At low $k$, its estimator loses power to fluctuations in the local mean density in the survey.  At high $k$, the small box power spectrum has a $1\%$ systematic offset due to differences
    in the simulation (Tree)-PM grid.   We remove both effects by scaling out this bias with
    Eq.~(\ref{eqn:debias}).}
    \label{fig:bias}
\end{figure}

In Fig.~\ref{fig:bias} we show the fractional difference in the mean power spectra.     At the lowest $k$, we clearly see the
effect of convolution by the window described by Eq.~(\ref{eqn:Pwindowed}).  
The convolution describes a net loss of power by aliasing into the $k=0$ mode.   At higher $k$, there remains a small $\sim 1\%$ bias.   {We have
verified that this is attributable to the difference in scaling of the (Tree)-PM grid noted above comparing small box simulations with the same seeds but with  $1/8$ scaling of the large box.}
Both types of bias effects are removed from the SSC 
considerations below by testing quantities scaled to the $\langle \hat P\rangle$ of the given set.  
More generally we can remove these effects by rescaling power spectrum
results in the subboxes by the mean bias
\begin{equation}
\hat P^\text{sub} \rightarrow
\frac{  P^\text{sm}}{  P^\text{sub}} \hat P^{\text{sub}},
\label{eqn:debias}
\end{equation}
where we have defined 
\begin{equation}
P^\text{X} = \frac{1}{N_s} \sum_{a=1}^{N_s}  \hat P^\text{X}_a
\end{equation}
as the average over the $N_s$ samples.

\begin{figure}[t]
    \centering
    \includegraphics[width=3.5in]{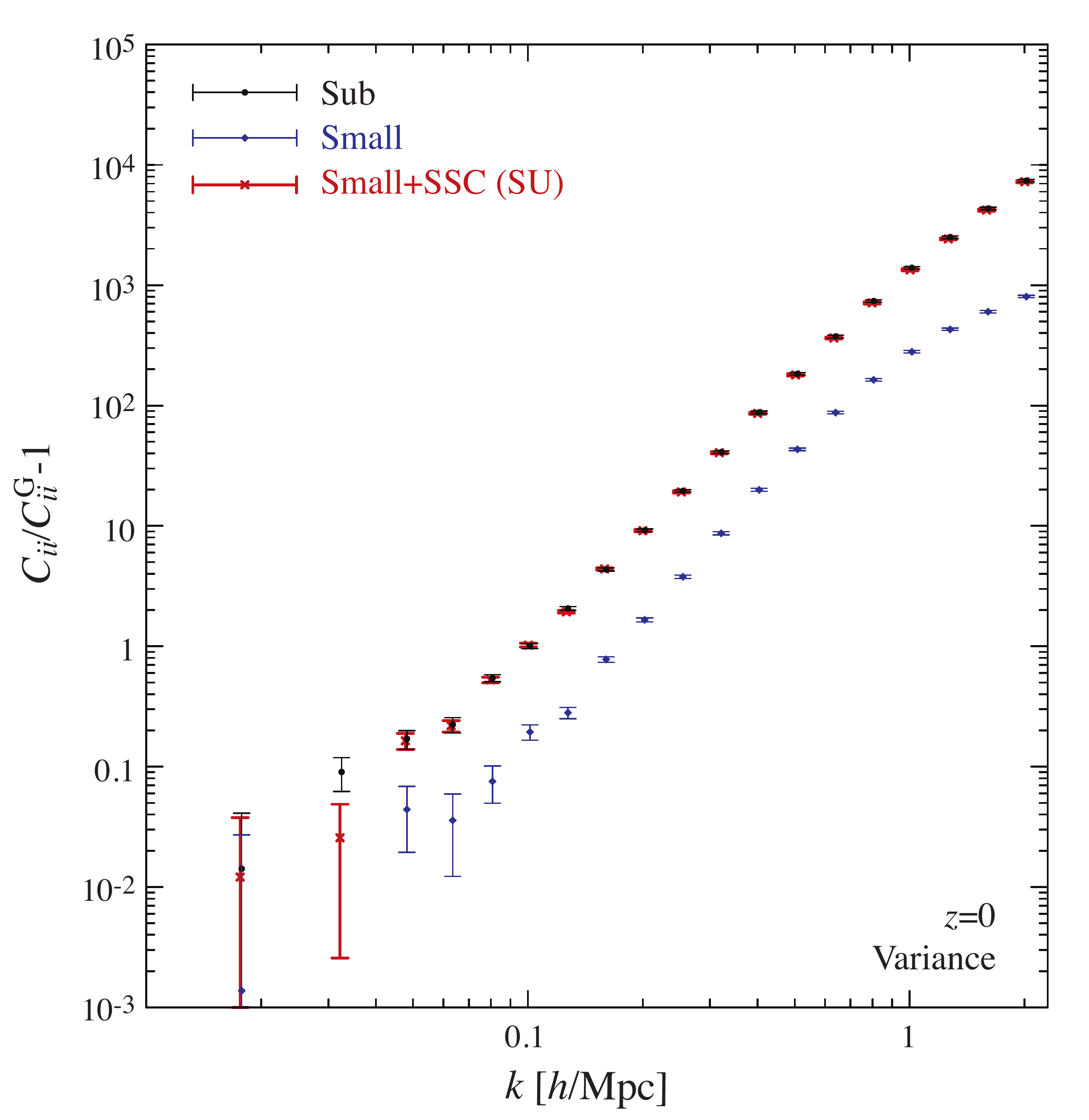}
    \caption{\footnotesize Power spectrum variance in excess over the Gaussian
    expectation $C_{ii}^\text{G}$ at $z=0$.   The SSC effect causes the subboxes of the large volume
    simulations to have up to an order of magnitude higher variance than found in small
    periodic boxes of the same volume.  Adding the SSC separate universe (SU) response
    to a background mode to the the small box variance models the effect to within the
    bootstrap errors of the simulation suites. Note that bootstrap errors between bins here and below are highly correlated. }
    \label{fig:variance}
\end{figure}

\begin{figure}[t]
    \centering
        \includegraphics[width=3.5in]{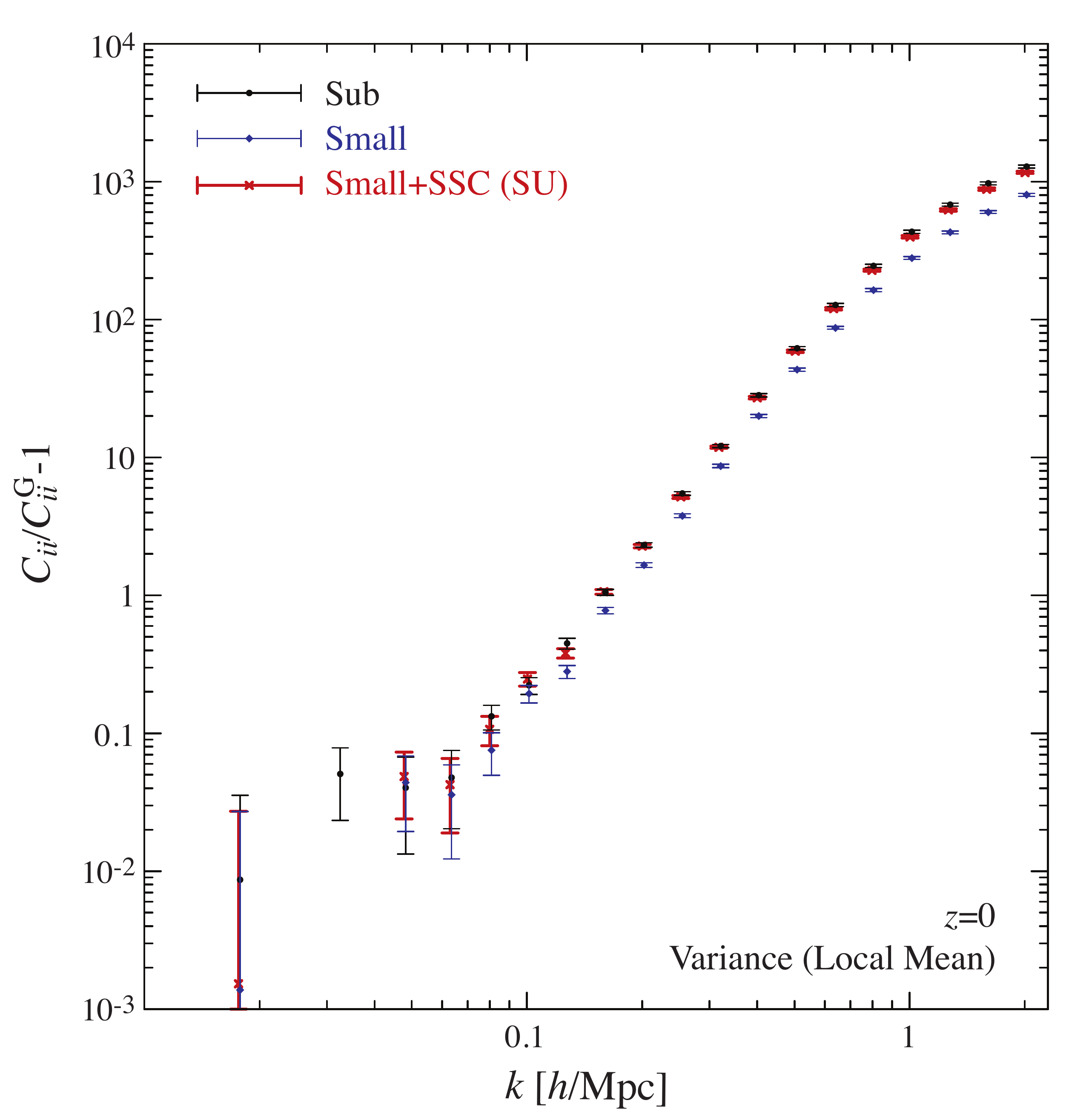}
    \caption{\footnotesize The same as in Fig.~\ref{fig:variance} but for power spectra with
    respect to the local mean density within the survey or subbox.  Here the SSC effect adds a comparable variance to other non-Gaussian effects and the total is modeled by the
    separate universe (SU) response to better than $10\%$ accuracy.}
    \label{fig:variancelocal}
\end{figure}

\begin{figure}[t]
    \centering
        \includegraphics[width=3.5in]{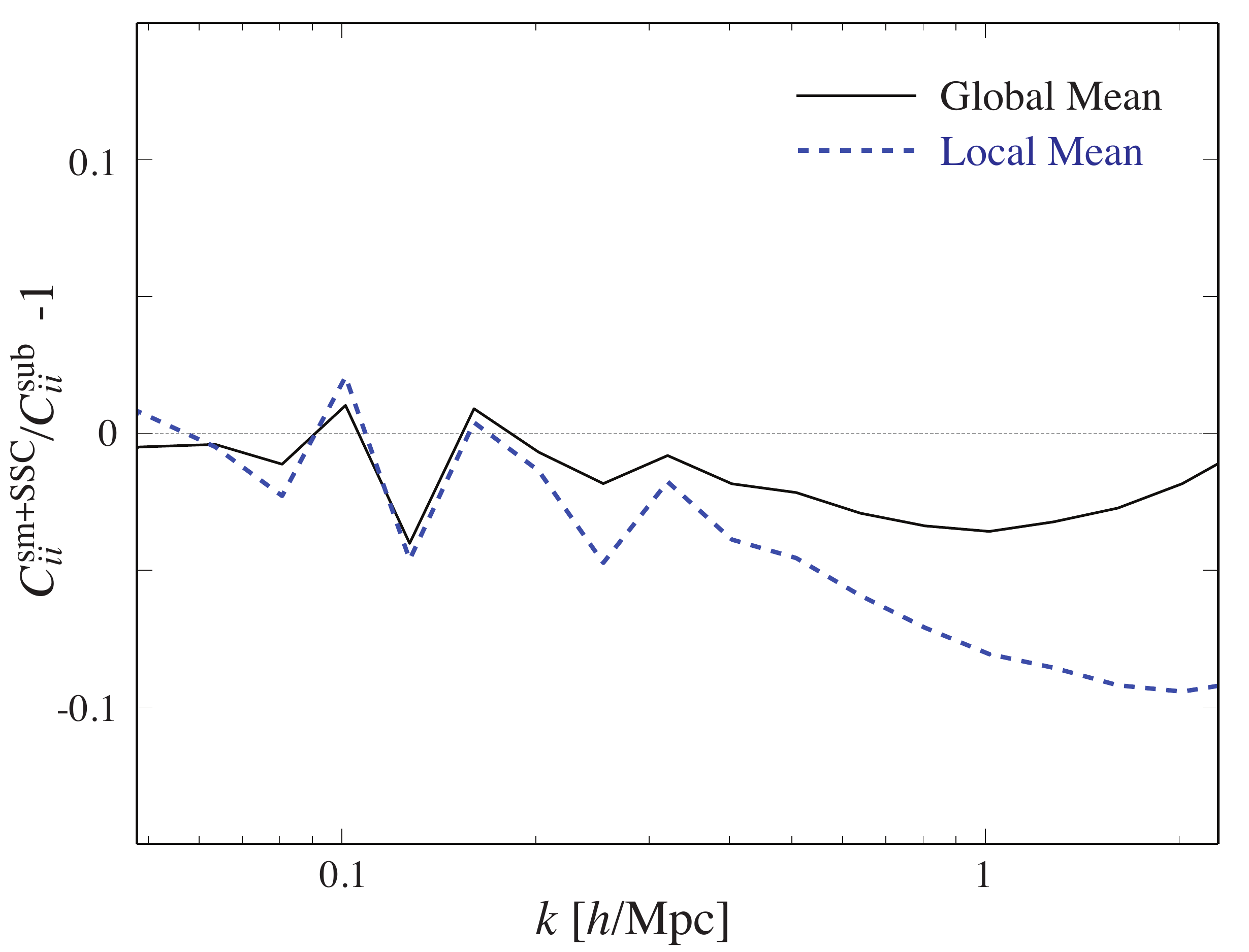}
        \caption{\footnotesize    Accuracy of the power spectrum variance model for the
        subbox covariance as the sum of small box and SSC variance effects. 
        For the global mean results of Fig.~\ref{fig:variance} the agreement is better
        than a few percent whereas for the local mean results of Fig.~\ref{fig:variance} it is
        better than 10\%.}
    \label{fig:varianceratio}
\end{figure}

\subsection{Power Spectrum Covariance}
\label{sub:super_sample_covariance}

For the covariance of the small box power spectra, we take the standard estimator 
\begin{eqnarray}
\hat C_{ij}^\text{sm} &=& \frac{N_s}{N_s-1} \Bigg[ \frac{\sum_{a=1}^{N_s} \hat P_a^\text{sm}(k_i)
\hat P_a^\text{sm}(k_j)}{N_s}\nonumber\\
 &&\hspace{5em} 
-  P^\text{sm}(k_i)P^\text{sm}(k_j)  \Bigg].
\end{eqnarray}
We then estimate the errors on the covariance through bootstrap resampling of the $N_s$ 
samples with replacement.

For the covariance of the subbox power spectra, we first form $s=1,...,N_s/N_l$ estimators from the $N_l=7$  fully independent large boxes
\begin{eqnarray}
\hat C_{ij}^s &=& \frac{N_l}{N_l-1} \Bigg[ \frac{\sum_{a=1}^{N_l} \hat P_{s a}^\text{sub}(k_i)
\hat P_{s a}^\text{sub}(k_j)}{N_l}\\&&
\hspace{4em}
 -  \frac{\sum_{a=1}^{N_l} 
\hat P_{s a}^\text{sub}(k_i) \sum_{b=1}^{N_l} \hat P_{s b}^\text{sub}(k_j)  } {N_l^2}  \Bigg]\nonumber
\end{eqnarray}
and then average the estimators
\begin{eqnarray}
\hat C_{ij}^\text{sub} = \frac{\sum_{s=1}^{N_s/N_l} \hat C_{ij}^s}{N_s/N_l}.
\end{eqnarray}
This construction assures that the end result is unbiased even if the subboxes in a 
given large box are themselves correlated, at the expense of slightly suboptimal errors.
For the error estimate, we bootstrap resample over the $N_s/N_l$ estimates.
We likewise form the estimator $\hat C_{ij}^W$ of the power spectrum referenced to
local means  out of $\hat P_W$ in the same way.

The SSC ansatz is that 
\begin{eqnarray}
\label{eqn:sscmodel}
C_{ij}^{\text{sub}} &=&
 C_{ij}^{\text{sm}} +  \sigma_b^2 P^\text{sm}(k_i) P^\text{sm}(k_j) \\
&& \times
    \frac{\partial\ln P^\text{SU}(k_i)}{\partial\delta_b} \frac{\partial\ln P^\text{SU}(k_j)}{\partial\delta_b}, \nonumber
\end{eqnarray}
for power spectra referenced to the global mean and
\begin{eqnarray}
\label{eqn:sscmodellocal}
C_{ij}^{W} &=&
 C_{ij}^{\text{sm}} +  \sigma_b^2 P^\text{sm}(k_i) P^\text{sm}(k_j) \\
&& \times \left(
    \frac{\partial\ln P^\text{SU}(k_i)}{\partial\delta_b}-2\right) \left( \frac{\partial\ln P^\text{SU}(k_j)}{\partial\delta_b}-2\right).\nonumber
\end{eqnarray}
We therefore construct these SSC model covariances from the small box estimates
and the separate universe response, denoted here with $P^\text{SU}$.
For the errors, we 
 bootstrap resample each
ingredient in the SSC ansatz (\ref{eqn:sscmodel}-\ref{eqn:sscmodellocal}):
$\hat P^\text{sm}$ from $N_s$ small box samples, $\partial \ln \hat P^\text{SU}/\partial \delta_b$
from $64$ separate universe response samples, and $\delta_b$ from $N_s$ subbox samples,
to combine them to get the error estimate on the covariance.
The dominant source of error on the mean
covariance is actually the draws of $\delta_b$ itself rather than the stochasticity of
the background response.

\begin{figure}[t]
    \centering
    \includegraphics[width=3.5in]{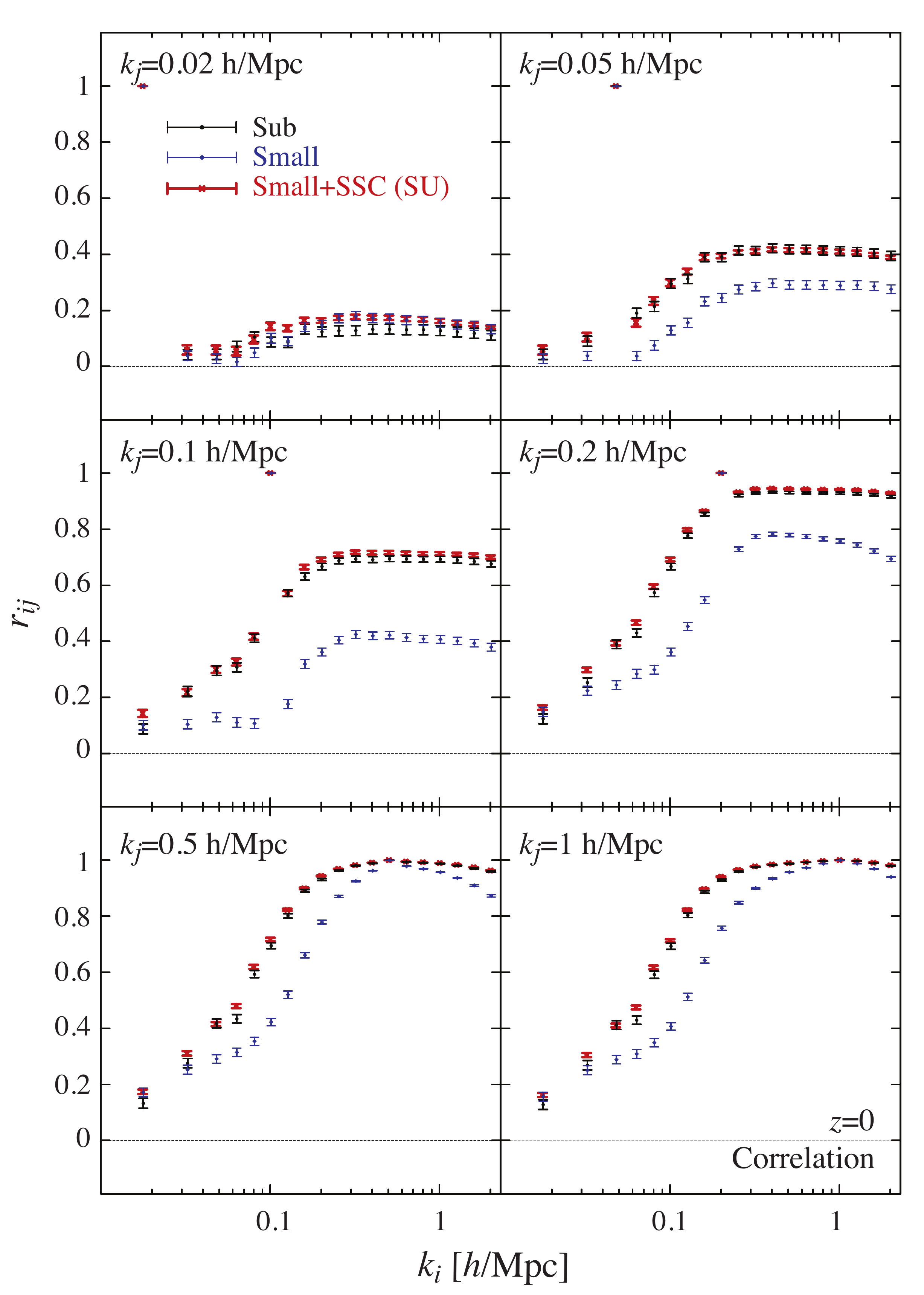}
    \caption{\footnotesize Power spectrum correlation coefficients between bins at $z=0$.   The SSC effect in the subboxes produces nearly fully correlated power spectrum changes or errors and is well-modeled by the separate universe response including the linear to highly 
    nonlinear correlation.     }
    \label{fig:correlation}
\end{figure}

\subsection{Results}
\label{sec:SSCresults}

In Fig.~\ref{fig:variance}, we show the  enhancement to power spectrum variances from non-Gaussian contributions
$C_{ii}/C^\text{G}_{ii}-1$ at $z=0$.  Here
$C_{ii}^\text{G}$ is the   Gaussian expectation of Eq.~(\ref{eqn:Cgaussian}) defined with 
{the mean} power spectra
from the simulation suites themselves.   Note that in these ratio statistics, any bias
due to convolution by the window shown in Fig.~\ref{fig:bias} drops out even before
debiasing through Eq.~(\ref{eqn:debias}).  

In the nonlinear regime $k \gtrsim 0.1 h/$Mpc, the non-Gaussian contributions
greatly exceed the Gaussian expectation.   Moreover the subbox covariance also exceeds the small box covariance, where super survey modes are absent,
by up to an order of magnitude indicating that the SSC effect is dominant.   This excess is modeled by the SSC ansatz to within the few percent bootstrap errors.

\begin{figure}[t]
    \centering
    \includegraphics[width=3.5in]{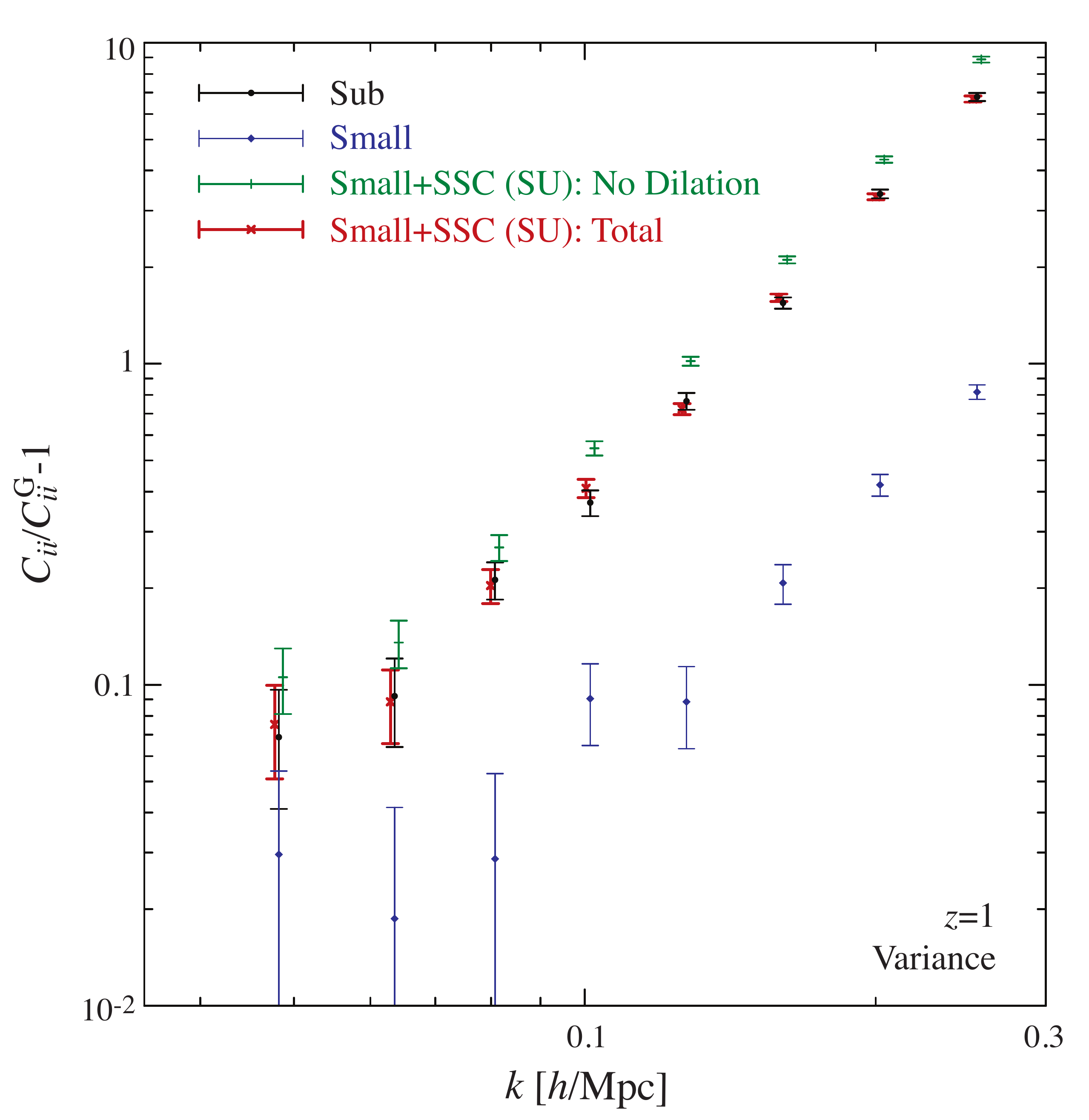}
    \caption{\footnotesize Power spectrum variance in excess over the Gaussian
    expectation $C_{ii}^\text{G}$ at $z=1$ highlighting the linear regime.    The separate universe
    total response to a background mode again models the SSC effect to within the bootstrap
    errors.   Here we also show that neglecting the dilation term in the response
    (``no dilation'') as in previous perturbative treatments in the literature produces poor agreement 
    even in the linear regime.   
    Agreement in the nonlinear regime and correlation coefficient is as good as at $z=0$.  }
    \label{fig:z1}
\end{figure}

We also show in Fig.~\ref{fig:variancelocal}, the same variance ratio but for estimated
power spectra referenced to the local mean $\hat P_W$ for the subboxes compared with 
the SSC model of Eq.~(\ref{eqn:sscmodellocal}).  
 For local mean results, the SSC effect is smaller and the
differences are  correspondingly larger but the total variance is still captured at the
$10\%$ level in a regime where it exceeds the Gaussian variance by $10^3$ (see Fig.~\ref{fig:varianceratio}).    Note that the bootstrap errors of all results are nearly fully correlated
in the nonlinear regime and so these difference may be consistent with statistical fluctuations
or differences in the non-SSC term.

In Fig.~\ref{fig:correlation}, we show the correlation coefficients 
\begin{equation}
r_{ij} \equiv \frac{C_{ij}}{\sqrt{C_{ii}C_{jj}}}
\end{equation}
for the global mean statistics, with bootstrap errors in {$C_{ij}$}.
Once the SSC effect dominates the variances in the nonlinear regime, the correlation coefficients of the subbox suite approach unity as expected for an effect that is determined by a single
template response.    This behavior is well-modeled by the SSC ansatz and differs qualitatively from the small box correlation coefficients.   The local mean correlation coefficients are equally well described by the SSC ansatz though the total is less
dominated by the SSC effect.   

In Fig.~\ref{fig:z1}
we show the variance ratio $C_{ii}/C^\text{G}_{ii}$ at $z=1$ to highlight effects in the
extended linear regime $k \lesssim 0.2-0.3 h/$Mpc.    The agreement between the subbox 
variance and the total SSC model is as good as at $z=0$ as is the correlation coefficients (not shown).  
Here we  also show the SSC
model where  the dilation term has been removed.    Even in the linear regime, 
the two models are clearly distinguished by the simulations.
This verifies that the linear dilation effect, omitted in previous studies, must be included
as part of the SSC model for accuracy in the quasilinear regime (cf.~\cite{dePutter:2011ah}).

{
\begin{figure}[t]
    \centering
    \includegraphics[width=3.5in]{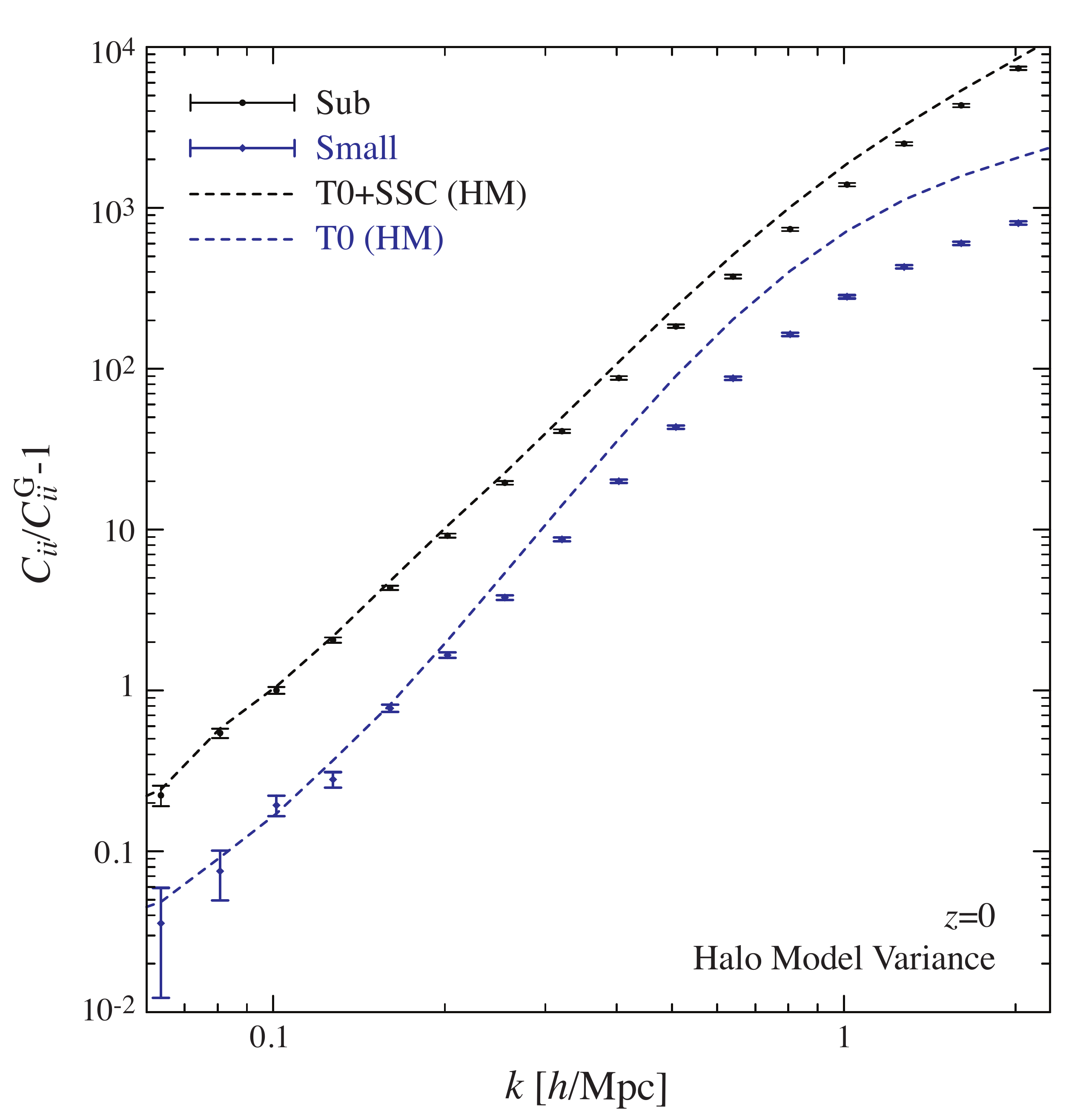}
    \caption{\footnotesize  Halo model for variance in excess over the Gaussian
    expectation $C_{ii}^\text{G}$ at $z=0$.   While the SSC contribution is well captured, the halo model overpredicts the variance of the small box simulations by up to a factor of 3.}
    \label{fig:halomodelcov}
\end{figure}
}

{
Finally in Fig.~\ref{fig:halomodelcov}, we compare the full halo model prescription for the
variance ratio from Eq.~(\ref{eq:covcon}) with the simulations.   In addition to the 
response function, whose accuracy was already tested in \S \ref{sec:sepres}, this includes
a model for the small box covariance $C_{ii}^{\text{T}0}$.   In the high-$k$, 1 halo regime,
the halo model for the latter differs  from the simulations  by up to 
 a factor of 3.  Note that very rare halos contribute substantially to this term and
that the ingredients of the halo model have not been as thoroughly tested in this regime.
On the other hand, the halo model for the SSC response term agrees nearly as well
as the separate universe response calibration in the difference between the small and subbox results. }

\section{Discussion}
\label{sec:discussion}

In this paper we have employed separate universe simulations to characterize accurately
all SSC effects from super-survey modes on matter power spectra measured from finite
volume surveys.   This approach automatically captures the separate effects of beat
coupling in the quasilinear regime, halo sample variance in the nonlinear regime and
a new dilation effect which changes scales in the power spectrum coherently across
the survey volume.   It accurately quantifies these effects with a handful of small
volume simulations once and for all, rather than the many thousands of survey specific volumes
required in the direct quantification.

Agreement between the SSC model, where the effect is described as the response
of the power spectrum to a change in the background density, and an extensive
suite of large volume simulations is excellent with no statistically significant deviations
within the domain of validity of the simulations $k \lesssim 3~ h/$Mpc for power spectra
referenced to the global mean.    
The SSC effect here provides the dominant non-Gaussian errors for a wide range 
of survey volumes and is encapsulated in a single response function that correlates
all modes in the spectrum.  
These results are relevant for the analysis of weak
lensing surveys where the global mean is defined by cosmological parameters.   

For power spectra referenced to the survey or local mean, relevant for galaxy surveys where the
mean density of tracers is typically estimated from the survey itself, the SSC effect is
a comparable effect to other non-Gaussian errors.   Here our calibration is still in good 
 agreement: better than the $10\%$ level even when the non-Gaussian variance
is a factor of $10^3$ larger than the Gaussian variance.  

We have also shown that the stochasticity of the response from volume to volume is small.
Hence to first approximation, these effects can be considered as an extra form of signal
rather than noise.   The unknown density fluctuation $\delta_b$ from super-survey modes
can be considered as a parameter that in the observed power spectrum which can be fit
for with the response template.    We have demonstrated that this should be an excellent
approximation in the quasilinear regime where the effect of dilation can bias precision measurements of BAO features in the power spectrum.

Agreement with the halo model for the power spectrum background response is also good implying that
a halo model description may allow for efficient extensions to 
alternate cosmologies.  In fact our separate universe analysis can also be used
to calibrate the elements of the halo model {and potentially improve agreement
for other non-Gaussian terms.     A crucial element of
the halo model is that halo bias is consistent with the mass function in the 
peak-background construction.}   Thus in principle halo bias can be calibrated from the
response of the halo mass function to a change in the background density
\cite{Schmidtetal:13}.   We intend to study such applications of the separate universe
simulations in a future work.

\smallskip{\em Acknowledgments.--}  
We thank M.\ Becker, M.\ Busha, B.\ Erickson,  G.\ Evrard, N.\ Gnedin,  A.\ Kravtsov,
D.\ Rudd,  R.\ Wechsler,  and the University of Chicago
Research Computing Center   for running, storing, and
allowing us to use the large-volume simulations in this study. 
{We also thank M.\ Becker, D.\ Rudd., N.\ Gnedin, A.\ Kravtsov and F.\ Schmidt for useful discussions.}
YL and WH  were supported
 by U.S.~Dept.\ of Energy
 contract DE-FG02-13ER41958
 and the
 Kavli Institute for Cosmological Physics at the University of
 Chicago through grants NSF PHY-0114422 and NSF PHY-0551142.   WH was additionally
 supported by the David and Lucile Packard Foundation. 
MT was supported by World
 Premier International Research Center Initiative (WPI Initiative),
 MEXT, Japan, by the FIRST program ``Subaru Measurements of Images and
 Redshifts (SuMIRe)'', CSTP, Japan, and by Grant-in-Aid for
 Scientific Research from the JSPS Promotion of Science (23340061).
 \vfill
\bibliography{sscsim}
\vfill
\end{document}